\documentclass[aps,prb,floats,twocolumn,showpacs,preprintnumbers,10pt]{revtex4}
\usepackage{natbib}
\usepackage{amsmath}
\usepackage{amsfonts}
\usepackage{amssymb}
\usepackage{graphicx}
\usepackage{dcolumn}
\usepackage{bm}
\usepackage{subfigure}
\usepackage{textcomp}
\usepackage{enumerate}

\begin{document}
\title{\textit{Hydrogen diffusion in the proton conductor Gd-doped barium cerate}}
\author{Jessica Hermet$^{1,2}$}
\author{Marc Torrent$^{1}$}  
\author{Fran\c{c}ois Bottin$^{1}$} 
\author{Guilhem Dezanneau$^{2}$} 
\author{Gr\'egory Geneste$^1$}
\email{gregory.geneste@cea.fr}
\affiliation{$^1$ CEA, DAM, DIF, F-91297 Arpajon, France \\
$^2$ Laboratoire Structures, Propri\'et\'es et Mod\'elisation des Solides, UMR CNRS 8580, \'Ecole Centrale Paris, Grande Voie des Vignes, 92295 Ch\^atenay-Malabry Cedex, France}
\date{\today}

\begin{abstract}
The energy landscape and diffusion barriers of protonic defects in Gd-doped BaCeO$_3$, a compound candidate as electrolyte for protonic ceramic fuel cells, have been investigated by density functional theory calculations, starting from a previously computed energy landscape consisting of 16 kinds of stable sites (8 close to dopants and 8 far from them). The simplified string method has been used to determine accurately the Minimum Energy Paths between those sites, that might imply either proton reorientations, intra-octahedral or inter-octahedral hopping mechanisms. At contrast with simple cubic perovskites such as barium stannate or barium zirconate, very different values for energy barriers (from 0.02~eV to 0.58~eV) are found in this highly distorted orthorhombic perovskite, and no specific process appears to be clearly rate-limiting. Some inter-octahedral hoppings (when possible) are found to be more favourable than the intra-octahedral ones, while reorientations exhibit a wide range of energy barriers.
\end{abstract}

\maketitle


\section{Introduction}

Since the discovery of protonic conductivity in aliovalent-doped SrCeO$_3$~\cite{Iwahara1981,Iwahara1983}, protonic conduction in perovskite-type oxides ABO$_3$ has been the subject of numerous studies, experimental as well as computational~\cite{Cherry1995,Iwahara1996,Kreuer1999,Kreuer2009,Norby2009}. The high protonic conductivity in perovskite oxides opens the way for a wide range of technological applications such as Protonic Ceramic Fuel Cells (PCFCs), hydrogen separators, etc.
However, if the diffusion of protons has been extensively explored by \textit{ab initio} calculations in cubic perovskites such as barium zirconate~\cite{Bjorketun2007,Sundell2007}, only very few works have studied this phenomenon in orthorhombic perovskites~\cite{Bilic2007}, although excellent proton conductors, such as SrCeO$_3$ or BaCeO$_3$, can be found among such systems.

Proton conductors are usually obtained by replacing some cations of a host oxide compound by cations with lower valence. In perovskite oxides having a tetravalent element on the B site (Ti, Zr, Ce, Sn), this can be done by inserting on this site a trivalent element. Such substitution creates charge-compensating oxygen vacancies that make the compound reactive with respect to water dissociation if it is put in contact with humid atmosphere. Such hydration reaction is commonly written, using Kr\"oger-Vink notations, as 
\begin{equation}\label{hydration}
 H_2O + V_O^{\bullet \bullet} + O_O^X \rightarrow 2 OH^{\bullet}_O.
\end{equation}
It generates protonic defects $OH^{\bullet}_O$, localized approximately along [100]-type directions inside the interoctahedral space of the perovskite network, and that can move from an oxygen site to another by simple thermal activation. Three possible motions of the proton in the perovskite network have been distinguished:

(i) the reorientation: the OH bond does not break and simply turns by $\approx$ 90$^{\circ}$ around the B-O-B axis containing the oxygen atom.

(ii) the intra-octahedral hopping: the proton leaves its oxygen site to move on another oxygen site of the same octahedron.

(iii) the inter-octahedral hopping: the proton leaves its oxygen site to move on another oxygen site that does not belong to the same octahedron.

In a previous work~\cite{Hermet2012}, we have studied by density-functional theory calculations the thermodynamics of hydration and oxidation of Gd-doped barium cerate BaCe$_{1-{\delta}}$Gd$_{\delta}$O$_{3-\frac{{\delta}}{2}}$ (BCGO). In particular, we have showed that hydration was an exothermic process and accurately determined the energy landscape of the proton near and far from the Gd dopant. We showed that this energy landscape can be well approximated by a surface with 16 kinds of local minima (8 in the close vicinity of the dopant, and 8 further). This complexity is the consequence of the highly distorted geometry of the host BaCeO$_3$, that adopts in its ground state the $Pnma$ space group. Consequently, proton migration throughout such energy surface involves many different energy barriers that need to be explored in order to get insight into proton conduction at the macroscopic scale. Previous works have studied proton migration in BaCeO$_3$, but only in the cubic phase~\cite{Munch1996,Kreuer1998,Munch2000}. Therefore, in this work, we present an exhaustive study of the Minimum Energy Paths associated to the possible motions for the proton in orthorhombic BCGO, and the values of their energy barriers.


\section{Computational details\label{details}}

\subsection{Method}
We have performed density functional theory (DFT) calculations using the plane-wave code ABINIT~\cite{Gonze2009,Bottin2008}.
The Generalized Gradient Approximation (GGA-PBE~\cite{Perdew1996}) was employed to describe electronic exchange and correlation. The calculations were carried out in the framework of the projector augmented wave (PAW) approach~\cite{Blochl1994,Torrent2008}. The same supercell as that of Ref.~\onlinecite{Hermet2012} was used: it consists of
80 atoms and has an orthorhombic symmetry ($Pnma$ space group). The First Brillouin Zone of this supercell was sampled by a 2$\times$2$\times$2 \textbf{k}-point grid, and the plane-wave cutoff was set to 20~Ha. The numerical accuracy on the total energies associated to this scheme is better than 1 mHa/atom. The cut-off radii of our PAW atomic data can be found in Ref.~\onlinecite{Hermet2012}.

In order to compute Minimum Energy Paths, the first task was to identify the stable sites of the proton in BCGO, which was previously achieved in Ref.~\onlinecite{Hermet2012}. This was performed by substituting in the 80-atom supercell one Ce by one Gd and introducing one hydrogen atom, that was placed in its different possible sites, close to the Gd dopant and far from it. In each configuration, the atomic positions were optimized until all the cartesian components of atomic forces were below 1$\times$10$^{-4}$~Ha/Bohr ($\approx$ 0.005~eV/{\AA}).

The possible energy barriers between pairs of stable protonic sites have then been computed using the so-called simplified string method~\cite{E2002,E2007}.
The simplified string method is an iterative algorithm allowing to find the Minimum Energy Path (MEP) between two stable configurations. It consists in discretizing the path into equidistant configurations, that we call "images". At each iteration, a two-step procedure is applied: first, each image is moved along the direction given by the atomic forces (evolution step), then the images are redistributed along the path in order to be kept equidistant (reparametrization step).
To determine the number of iterations of string method, we used an optimization criterion related to the energy of the images: the optimization of the MEP is stopped when the total energy (averaged over all the images) difference between an iteration and the previous one is lower than 1$\times$10$^{-5}$~Ha. In such an algorithm, the result should be carefully converged with respect to the number of images along the path, which forced us to use up to 19 images in the case of some intra-octahedral hopping processes. Once the MEP has been correctly converged, the maximum energy along the path provides us the transition state, and thus the energy barrier of the corresponding process (hopping or reorientation). Finally, we point out that all the atoms of the supercell were allowed to move during the computation of the MEP, thus providing energy barriers in a ``fully-relaxed'' system.

For the sake of numerical efficiency, we have used the three traditional levels of parallelization present in the ABINIT code (\textbf{k}-points, bands, plane waves) together with a fourth level on the images of the system used to discretize the MEP. This fourth level has a quasi-linear scalability and, since the number of images used to discretize the path can be as large as 19, thousands of cpu cores can be used to compute and relax the MEP with high efficiency. Typical jobs were done on 3000~cpu cores using these four parallelization levels, allowing us to take maximal benefit of the potentialities of parallel supercomputers.

\subsection{Approximations to the computation of energy barriers}
Additional remarks have to be mentionned about the limitations of our approach and the approximations made to compute the energy barriers.

First of all, the string method, like the Nudged Elastic Band method, allows to compute the Minimum Energy Paths between two stable configurations and thus to obtain ``fully-relaxed'' (static) barriers, as opposed to ``dynamical'' barriers that would be obtained, for instance, by counting the occurences of each event within a molecular dynamics run and fitting the rates by an Arrhenius law. Static barriers neglect some collective effects and the so-called recrossing processes. In theory, they make sense only if the whole structure is able to relax instantaneously when the proton moves from a stable position to another. However, the time scale associated to the hydrogen motion is much smaller than the ones of the deformation of the surrounding structure, which involves much softer phonon modes. The motion of protons in an unrelaxed envionment would naturally lead to higher barriers than those calculated from fully-relaxed DFT calculations. Nevertheless, as shown by Li and Wahnstr{\"o}m~\cite{Li1992} in metallic palladium, the jump of the proton has to be considered in a reverse way. Due to the vibrations of surrounding atoms and to the high vibration frequency of hydrogen, protons currently jump at a moment where the surrounding atoms are in a geometrical configuration close to the calculated relaxed one. That is why the calculated barriers can result very close to those currently observed. Further work should be nervertheless necessary to verify that the proton jump, for instance during {\it ab inito} Molecular Dynamics simulations, occurs for a geometry of surrounding atoms close to that calculated  in the fully relaxed DFT static scheme.

Second, the present barriers do not include quantum contributions from zero-point motions. They are valid in the limit where nuclei can be considered as classical particles. If this approximation is correct for heavy atoms in the temperature range interesting PCFCs, this is not so obvious for the proton~\cite{Sundell2007}. Indeed, proton tunnelling might occur and thus significantly lower the barrier height, especially in the hopping case~\cite{Zhang2008}. This approximation leads to overestimated barriers.

Last, the use of the Generalized Gradient Approximation tends to underestimate the activation energy for proton transfer in hydrogen-bonded systems~\cite{Bjorketun2007}. This underestimation is due to an over-stabilization of structures in which an hydrogen is equally shared between two electronegative atoms~\cite{Barone1996}.

Consequently, the barriers presented in this work purely reflect the GGA potential energy surface of the proton in its host compound. They are static barriers, free from collective, dynamical and quantum effects.


\section{Review of preliminary results: structure of BaCeO$_3$ and protonic sites}
\setcounter{subsubsection}{0}

\subsection{BaCeO$_{3}$ and BCGO structure}

As many perovskites~\cite{Goudochnikov2007}, BaCeO$_3$ has an orthorhombic structure ($Pnma$ space group~\cite{Knight2001}) at room temperature (RT). At high temperature, it undergoes three structural phase transitions, the first one at $\approx$ 550 K towards an $Imma$ structure, and the second one at $\approx$ 670 K towards a rhombohedral $R\bar{3}c$ structure. At very high temperature ($\approx$ 1170 K), it eventually evolves towards the parent $Pm\bar{3}m$ cubic structure, that of the ideal perovskite. The presence of dopants randomly distributed throughout the matrix may change transition temperatures. 
However Melekh {\it et al.}~\cite{Melekh1997} found that the first transition in 10\%-Gd-doped BaCeO$_3$ occurs around 480-540~K, close to the one they found for pure BaCeO$_3$ of 533~K. At RT, Gd-doped BaCeO$_3$ is therefore orthorhombic.

Our calculations provide optimized configurations and Minimum Energy Paths. These computations are thus relevant when performed in combination with the ground state structure of BaCeO$_3$, {\it i.e.} the orthorhombic $Pnma$ structure, which was used as starting point in all the calculations, and globally preserved along the optimizations procedures. The computed energy barriers can therefore be used to understand proton diffusion in BCGO below $\approx$ 550 K. However, from a more general point of view, the present results provide a useful microscopic insight into proton diffusion in a low-symmetry perovskite compound, typical of those used as electrolytes in Proton Ceramic Fuel Cells (the $Pnma$ structure is common to many perovskites such as cerates, zirconates, titanates or stannates).

The structural parameters obtained for BaCeO$_{3}$ and BCGO within the present scheme can be found in Ref.~\onlinecite{Hermet2012}. They are in excellent agreement with experiments, despite a slight overestimation of the lattice constants related to the use of the GGA.

\subsection{Protonic sites in perovskites: general considerations\label{stable-site}}
As previously explained, proton conduction in an ABO$_3$ perovskite compound -- where B is a tetravalent element -- might be obtained by substituting B atoms by trivalent elements such as Gd (this creates oxygen vacancies by charge compensation) and by subsequently exposing the new compound to humid atmosphere. The protons as charge carriers then appear through the dissociation of water molecules into the oxygen vacancies, according to the well-known hydration reaction (see equation~\ref{hydration}).

The precise location of the stable protonic sites in the perovskite network seems to strongly depend on the lattice parameter and distortion of the host compound.
It is commonly admitted that protons are bonded to an oxygen atom and remain in the form of hydroxyl groups located on oxygen sites. But the orientation of the O-H bond is not that clear. On the one hand, it was proposed that it could be oriented along the BO$_6$ octahedra edge because of its dipolar moment~\cite{Kreuer1995,Kreuer2009}, leading to 8 possible sites per oxygen atom. On the other hand, previous experimental~\cite{Hempelmann1998} and ab initio~\cite{Glockner1999,Davies1999,Tauer2011} studies have found only four sites per oxygen atom oriented along the pseudo-cubic directions. 

In fact, the stable protonic sites seem to be indeed 

(i) along or close to the octahedra edge for perovskites with relatively small lattice constant $a_0$, such as SrTiO$_3$~\cite{Cherry1995,Matsushita1999} or LaMnO$_3$~\cite{Cherry1995} ($a_0$=3.91~\AA), leading to the existence of 8 protonic sites per oxygen atom, 

(ii) along the pseudo-cubic directions for perovskites with large lattice constant, such as SrZrO$_3$~\cite{Davies1999} or BaCeO$_3$~\cite{Glockner1999,Tauer2011,Hermet2012} (pseudo-cubic lattice constant $a_0$=4.14 and 4.41~\AA{} respectively), leading to the existence of 4 protonic sites per oxygen atom.

This trend can easily be explained : as the lattice constant decreases, the nearest oxygen gets closer and closer to the proton, attracting it sufficiently (through hydrogen bond) to bend the O-H bond towards the octahedron edge.

\subsection{Protonic sites in Gd-doped BaCeO$_3$}

\begin{figure}[h]
 \includegraphics[scale=0.32]{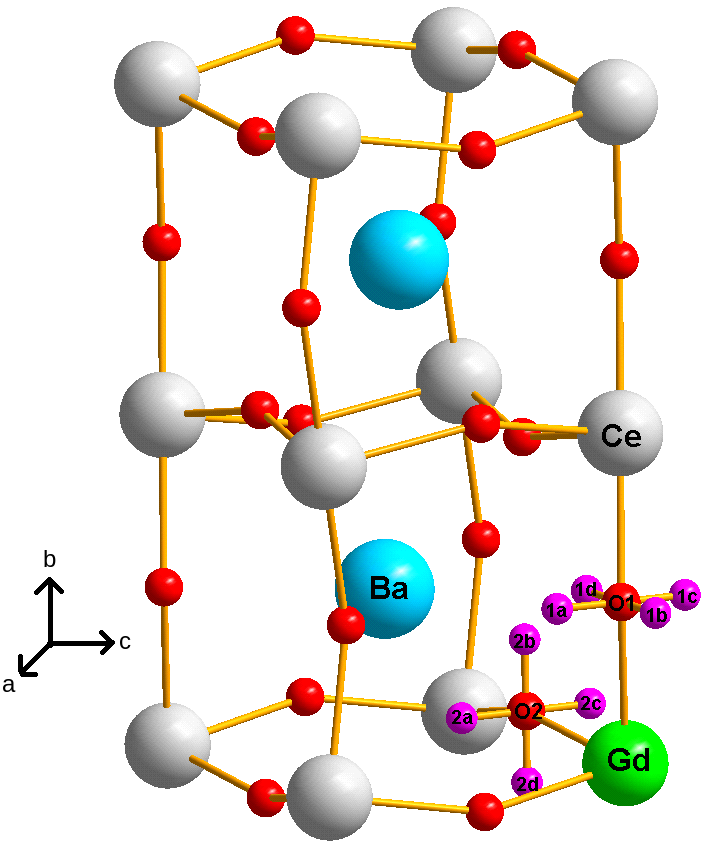}
  \caption{The 8 stable positions for the proton around the Gd dopant.\label{8protons}}
\end{figure}

In our previous calculations on BCGO~\cite{Hermet2012}, which has a large pseudo-cubic lattice constant of 4.41~\AA, we found indeed four stable protonic sites per oxygen atom. Considering that the $Pnma$ structure contains two inequivalent oxygen atoms O1 and O2, this leads to the existence of 8 inequivalent stable positions for the proton, if we ignore the symmetry-breaking caused by the presence of dopants. These positions have been labeled 1a, 1b, 1c and 1d for those attached to O1 (apical oxygen), and 2a, 2b, 2c and 2d for those attached to O2 (equatorial oxygen), see Fig.~\ref{8protons}.

However, when one Ce atom is replaced by a Gd dopant, both the translational symmetry and the symmetry between the four equatorial oxygens O2 of the first coordination shell of this specific B site are broken. More precisely, the presence of Gd splits the four O2 into two pairs of symmetry equivalent oxygen atoms (O2 and O2'). The four inequivalent protonic sites related to O2 (2a, 2b, 2c and 2d) are thus split into 8 inequivalent sites, called 2a, 2b, 2c, 2d, 2a', 2b', 2c' and 2d'. The first coordination shell of Gd exhibits therefore 12 kinds of inequivalent protonic sites. Beyond this shell, the symmetry-breaking is even more complex.

Nevertheless, we have shown in Ref.~\onlinecite{Hermet2012} that this new emerging complex protonic energy landscape can be very well approximated by a surface containing 16 kinds of inequivalent local minima: 8 corresponding to the 8 sites shown in Fig.~\ref{8protons} close to a Gd dopant, and 8 associated to the same sites "far" from the dopant, {\it i.e.} beyond its first oxygen coordination shell. Tab.~\ref{energy-proton} gives the relative energy associated to each site (taken from Ref.~\onlinecite{Hermet2012}): in the first coordination shell of Gd, only 8 sites among the 12 can be considered as non-equivalent. Beyond also, only the same 8 kinds of sites can be considered as non-equivalent with a very good accuracy. In other words, the symmetry-breaking caused by the presence of dopants can be considered as having no significant influence on the energy landscape of the protonic defects. In order to distinguish the sites of these two families, we introduce another letter, "n" (for a site \textbf{near} the dopant), or "f" (for a site \textbf{far} from the dopant).

To summarize, the 16 kinds of stable positions are labeled by 
\begin{itemize}
 \item a number (1 or 2) corresponding to the oxygen type (apical and equatorial, respectively),
 \item a letter (``a", ``b", ``c" or ``d") corresponding to the O-H direction (shown in figure~\ref{8protons}),
 \item and another letter, ``n" for a site \textbf{near} the dopant, or ''f" for a site \textbf{far} from the dopant.
\end{itemize}

\begin{table}
 \centering
 \begin{tabular}{lp{2cm}lp{2cm}}
     & Gd-OH-Ce & &Ce-OH-Ce \\
 1an & 0.00 & 1af &0.09 \\
 1bn & 0.01 & 1bf &0.08 \\
 1cn & 0.11 & 1cf &0.25 \\
 1dn & 0.00 & 1df &0.14 \\
 2an (2a'n) & 0.17 (0.16) & 2af &0.25 \\
 2bn (2b'n) & 0.05 (0.05) & 2bf &0.12 \\
 2cn (2c'n) & 0.15 (0.13) & 2cf &0.29 \\
 2dn (2d'n) & 0.08 (0.09) & 2df &0.23 \\
 \end{tabular}
 \caption{Energy (in eV) of the possible proton binding sites in BCGO relative to the most stable one (1an).}
\label{energy-proton}
\end{table}

In the presence of a dopant, the OH bond might slightly deviate from the pseudo-cubic direction: usually the proton is expected to bend towards the dopant due to the opposite formal charge of the corresponding defects (+1 for the protonic defect $OH_O^{\bullet}$ versus -1 for the dopant defect $Gd_{Ce}^{'}$). But it also depends on the dopant size~\cite{Davies1999,Bjorketun2007}.

In the present case, the proton has indeed a tendency to bend slightly towards the dopant, but with a deviation from the pseudo-cubic direction lower than 10\textdegree{} (see Tab.~\ref{angle-values}). 
It is possible to divide the eight stable sites into two categories: either the proton is able to hop from one octahedron to another (a/b-type) or not (c/d-type). The c/d-type site shows a noticeable bending (around 5\textdegree) while the a/b-type are almost perfectly aligned along the pseudo-cubic direction. This absence of bending may be due to the stabilization of a/b-type sites by an hydrogen bond with the facing oxygen, which is in those cases rather close. This hydrogen bond would be dominant over the proton-dopant interaction, especially since the dopant is much further than for a c/d-type site.

\begin{figure}
\centering
 \includegraphics[scale=0.32]{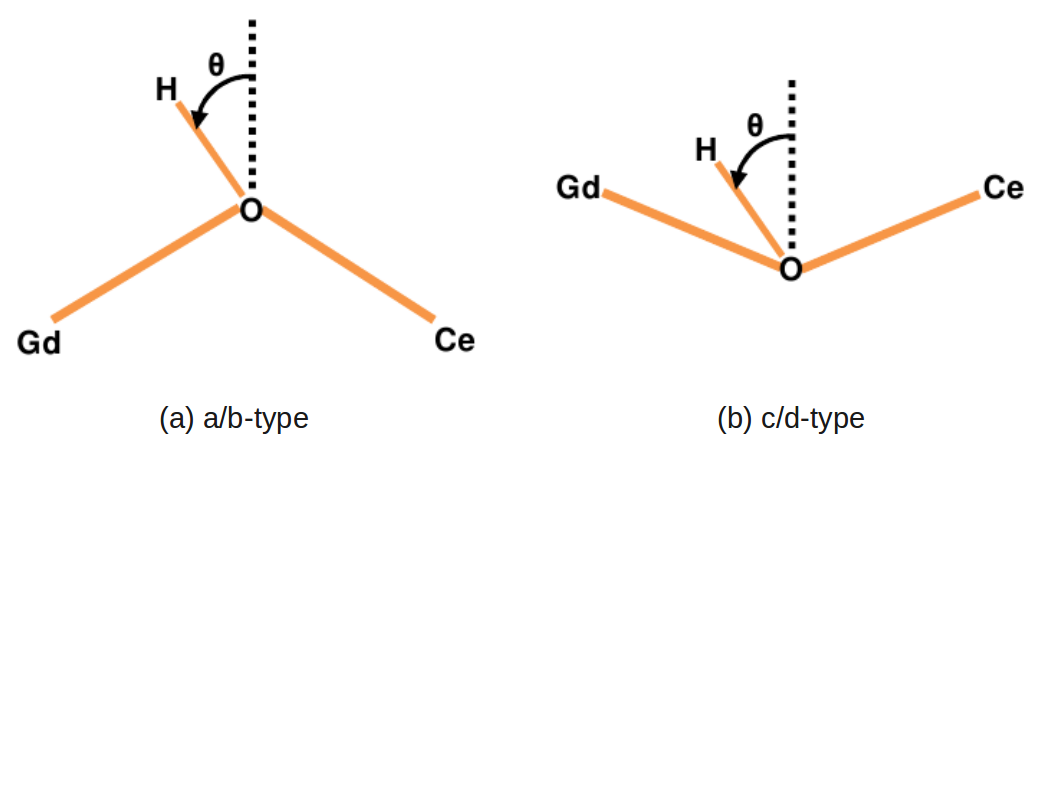}
\caption{Angles between the pseudo-cubic direction and the actual O-H bond for the two subcategories of protonic sites.}
\label{angle-picture}
\end{figure}


\begin{table}
 \centering
 \begin{tabular}{cdd}
  Position & \multicolumn{1}{l}{$\theta$ near Gd} & \multicolumn{1}{l}{$\theta$ far from Gd}\\
  1a & -0.1$\textdegree$ & 0.2$\textdegree$ \\
  1b & -0.1$\textdegree$ & 0.2$\textdegree$ \\
  1c &  5.3$\textdegree$ & 0.5$\textdegree$ \\
  1d &  3.5$\textdegree$ & 0.5$\textdegree$ \\
  2a & -0.5$\textdegree$ & 0.6$\textdegree$ \\
  2b &  1.6$\textdegree$ & 1.2$\textdegree$ \\
  2c &  5.0$\textdegree$ & 2.1$\textdegree$ \\
  2d &  8.2$\textdegree$ & 4.9$\textdegree$ \\
 \end{tabular}
\caption{Values of the angle described in figure~\ref{angle-picture}, for a proton near a dopant, and far from a dopant.}
\label{angle-values}
\end{table}

Note that in perovskites with smaller lattice constant, the bending is usually stronger, but also highly dopant-dependent. Bjorketun {\it et al.}\cite{Bjorketun2007} have studied this dependence in BaZrO$_3$ and got a bending angle from 6.9\textdegree{} for Gadolinium up to 20.4\textdegree{} for Gallium. An even higher bending of around 30\textdegree{} for Scandium, Yttrium or Ytterbium have been found in SrZrO$_3$~\cite{Davies1999}.

\section{Energy barriers}
\setcounter{subsubsection}{0}

We have seen that the energy landscape of stable protonic sites in Gd-doped BaCeO$_3$ is really complex, due to the distortions of the $Pnma$ structure and the presence of dopants. As a result, there are many different values for the energy barriers, associated to several diffusion mechanisms, even by considering the simplified energy landscape (with 16 minima) presented previously.

\subsection{The three different mechanisms: reorientation, intra-octahedral and inter-octahedral hopping}

In an ideal cubic perovskite, there are two kinds of processes for the proton motion: reorientation and transfer (or hopping)~\cite{Bjorketun2005}, to which only two different energy barriers can be associated, provided the proton is assumed to be far from any dopant. In BaZrO$_3$, the reorientation (resp. transfer) barrier is 0.14~eV (resp. 0.25~eV), while in cubic BaTiO$_3$~\cite{Gomez2005}, it is 0.19~eV (resp. 0.25~eV). In such simple systems, each proton in a stable site has four different possibilities to move: two reorientations and two intra-octahedral hopping, the inter-octahedral hopping being considered as unlikely (because the oxygen facing the OH group is too far).

\begin{figure}
\centering
\includegraphics[scale=0.45]{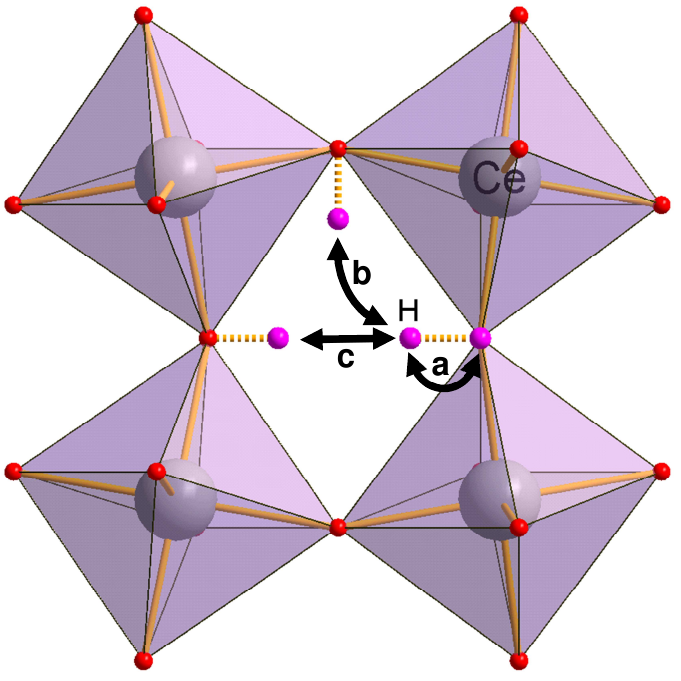}
\caption{Possible motions of the proton in the perovskite $Pnma$ structure. (a) reorientation, (b) intra-octahedral hopping, (c) inter-octahedral hopping.}
\label{3motions}
\end{figure}

However, the existence of tilts of oxygen octahedra, very common in perovskite oxides~\cite{Goudochnikov2007} having low tolerance factor $t=\frac{r_A+r_O}{\sqrt{2}(r_B+r_O)}$, makes the inter-octahedral hopping more likely in these strongly distorted structures (Fig.~\ref{3motions}), because some inter-octahedral oxygen-oxygen distances might be considerably lowered by the antiferrodistortive motions of the oxygen atoms. The proton may thus jump directly from one octahedron to another (one inter-octahedral hopping instead of two intra-octahedral hoppings), which might result in an increase of the macroscopic diffusion coefficients. Tab.~\ref{flip-inter} emphasizes the link between the tolerance factor $t$, the perovskite structure, and the possibility of inter-octahedral transfer according to the works mentionned.

As explained in Sec.~\ref{stable-site}, in perovskites with small lattice constant ($\leq$ 4.0 \AA), the proton in its stable site tends to bend towards one oxygen atom of one neighboring octahedron instead of being equidistant from both neighboring oxygens. In such systems, there are therefore twice more stable sites than in perovskites with larger lattice constant, so that an additional rotational mechanism might exist, corresponding to the slight reorientation of OH, bending from the edge of one neighboring octahedron to the other. This mechanism was previously called ``flip"~\cite{Rasim2011} or ``bending"~\cite{Munch1996} or ``inter-octahedron hopping"~\cite{Matsushita1999} (but ``inter-octahedron reorientation" should be less confusing, since the bond between O and H is not broken during this process). 
However, the energy barrier of the flip is usually rather low ($\lesssim 0.1$~eV~\cite{Matsushita1999}), and thus most of the time neglected.
It can also be seen as part of the intra-octahedral transfer mechanism : before jumping from one oxygen to another, there is a little reorientation of the proton in order to get an alignment O-H...O.  The intra-octahedral transfer would thus be a two-step mechanism with bending then stretching.

Tab.~\ref{flip-inter} illustrates the possible correlation between the lattice parameter and the possibility to flip for several proton conductor perovskites.  
Note that some studies found a possible inter-octahedral transfer in small cubic perovskite such as SrTiO$_3$~\cite{Kreuer1999,Munch1999a,Munch2000} or even in cubic perovskites with large lattice constant such as BaZrO$_3$~\cite{Merinov2009}, in contradiction with other works~\cite{Munch2000,Gomez2005}.

\begin{table}
\centering
\begin{tabular}{lccccc}
Perovskite & a$_0$ (\AA)& t & Structure & Flip & Inter\\
SrCeO$_3$~\cite{Munch1999}                   & 4.29 & 0.89 & $Pnma$ & no & yes\\
CaZrO$_3$~\cite{Davies1999,Islam2001,Shi2005,Gomez2005} & 4.04 & 0.92 & $Pnma$ & no & yes \\
BaCeO$_3$~\cite{Munch1999}                   & 4.41 & 0.94 & $Pnma$ & no  & yes \\
SrZrO$_3$~\cite{Davies1999,Shi2005,Gomez2007}& 4.14 & 0.95 & $Pnma$ & no  & yes \\
CaTiO$_3$~\cite{Munch1999a,Munch2000,Gomez2005} & 3.85 & 0.97 & $Pnma$ & yes & yes \\
BaZrO$_3$~\cite{Shi2005,Gomez2005,Bjorketun2005} & 4.25 & 1.01 & $Pm\bar{3}m$ & no  & no \\
SrTiO$_3$~\cite{Munch1999a,Munch2000}         & 3.91 & 1.01 & $Pm\bar{3}m$ ($I4/mcm$) & yes & no  \\
BaSnO$_3$~\cite{Bevillon2008a}               & 4.16 & 1.03 & $Pm\bar{3}m$ & no  & no  \\
BaTiO$_3$~\cite{Gomez2005}                    & 4.06 & 1.07 & $Pm\bar{3}m$ ($R3m$) & no & no  \\
\end{tabular}
\caption{Pseudo-cubic lattice constant a$_0$ from DFT calculations (GGA), tolerance factor $t$ calculated from Shannon ionic radii, crystal space group of different perovskite oxides, and whether flip or inter-octahedral hopping can occur or not. For BaTiO$_3$, the high-temperature cubic structure is considered, which is the one simulated in Ref.~\onlinecite{Gomez2005}. The cubic structure is also considered for SrTiO$_3$, rather than the low-temperature tetragonal structure. In those two cases, the ground state space group is given between parenthesis.}
\label{flip-inter}
\end{table}

\subsection{Energy barriers and Minimum Energy Paths}
 
Using the string method, the Minimum Energy Paths joining the various stable sites have been computed, giving access to the transition states and thus the energy barrier for the corresponding proton motion. These energy barriers are provided in Tab.~\ref{barrier-value}. Note that the barriers far from dopants ({\it i.e.} from "f" to "f") have been also computed in a 80-atom supercell without dopant and a +1 charge state (to simulate the protonic defect), compensated by a uniform charged background. The energy barrier values obtained are identical to the ones obtained in the doped supercell within 0.01 eV and are presented in the Appendix.

\begin{table}
 \centering
\begin{tabular}{c cc cc cc cc cc}
 &\multicolumn{4}{c}{Reorientation} & \multicolumn{4}{c}{Intra} &\multicolumn{2}{c}{Inter} \\
From & To &\emph{{$\Delta$E}} & To & \emph{{$\Delta$E}} & To & \emph{{$\Delta$E}}  & To & \emph{{$\Delta$E}} & To & \emph{{$\Delta$E}} \\
\hline
1an & 1bn & 0.50 & 1dn & 0.10 & 2dn & 0.37 & 2df & 0.58 & 1bf & 0.24 \\
1bn & 1cn & 0.30 & 1an & 0.49 & 2dn & 0.32 & 2df & 0.48 & 1af & 0.24 \\
1cn & 1dn & 0.05 & 1bn & 0.20 & 2bn & 0.29 & 2bf & 0.43 &     &      \\
1dn & 1an & 0.09 & 1cn & 0.16 & 2bn & 0.36 & 2bf & 0.52 &     &      \\
2an & 2bn & 0.31 & 2dn & 0.15 & 2cn & 0.22 & 2cf & 0.40 & 2af & 0.25 \\
2bn & 2cn & 0.28 & 2an & 0.43 & 1cn & 0.35 & 1cf & 0.51 & 2bf & 0.21 \\
    &	  &	 &     &      & 1dn & 0.31 & 1df & 0.47 &     &      \\
2cn & 2dn & 0.03 & 2bn & 0.18 & 2an & 0.23 & 2af & 0.45 &     &      \\
2dn & 2an & 0.23 & 2cn & 0.09 & 1an & 0.29 & 1af & 0.44 &     &      \\
    &	  &	 &     &      & 1bn & 0.24 & 1bf & 0.39 &     &      \\
\hline		        	   	                     	     
1af & 1bf & 0.54 & 1df & 0.14 & 2df & 0.50 & 2dn & 0.44 & 1bf & 0.19 \\
    &	  &	 &     &      &     &	   &     &	& 1bn & 0.16 \\
1bf & 1cf & 0.33 & 1af & 0.54 & 2df & 0.45 & 2dn & 0.40 & 1af & 0.20 \\
    &	  &	 &     &      &     &	   &     &	& 1an & 0.16 \\
1cf & 1df & 0.06 & 1bf & 0.18 & 2bf & 0.36 & 2bn & 0.32 &     &      \\
1df & 1af & 0.08 & 1cf & 0.15 & 2bf & 0.42 & 2bn & 0.39 &     &      \\
2af & 2bf & 0.36 & 2df & 0.17 & 2cf & 0.39 & 2cn & 0.36 & 2af & 0.21 \\
    &	  &	 &     &      &     &	   &     &	& 2an & 0.17 \\
2bf & 2cf & 0.33 & 2af & 0.49 & 1cf & 0.47 & 1cn & 0.42 & 2bf & 0.16 \\
    &	  &	 &     &      & 1df & 0.44 & 1dn & 0.39 & 2bn & 0.13 \\
2cf & 2df & 0.02 & 2bf & 0.17 & 2af & 0.36 & 2an & 0.28 &	&      \\
2df & 2af & 0.20 & 2cf & 0.08 & 1af & 0.37 & 1an & 0.34 &	&      \\
    &     &      &     &      & 1bf & 0.31 & 1bn & 0.28 &     &      \\
\end{tabular}
\caption{Energy barriers (eV) for proton reorientation, intra-octahedral hopping ("intra") and inter-octahedral hopping ("inter").}
\label{barrier-value}
\end{table}

Starting from a given initial position, the possible motions for the proton are: two reorientations, two intra-octahedral hopping, and possibly one inter-octahedral hopping if the configuration is favorable (which is the case for a and b-type positions where the oxygen atom facing the proton is close enough). Looking at Tab.~\ref{barrier-value}, we can notice that barriers between two ``near'' sites or two ``far'' sites, corresponding to reorientation barriers, are very similar (difference within 0.05 eV). This is expected as the energy surface of protons bonded to an oxygen 1st neighbor of a dopant is almost simply shifted by 0.1~eV compared to that of protons far from the dopant, leading to similar energy landscape. However, the case of hopping is more complicated since the Coulomb interaction between H and Gd prevents hydrogen from easily escaping from the dopant neighborhood. Thus, hopping barriers between a ``near'' site and a ``far'' site have usually a higher value that the ones corresponding to the backward motion.

\begin{center}
\begin{figure*}
\centering
\includegraphics[scale=0.625]{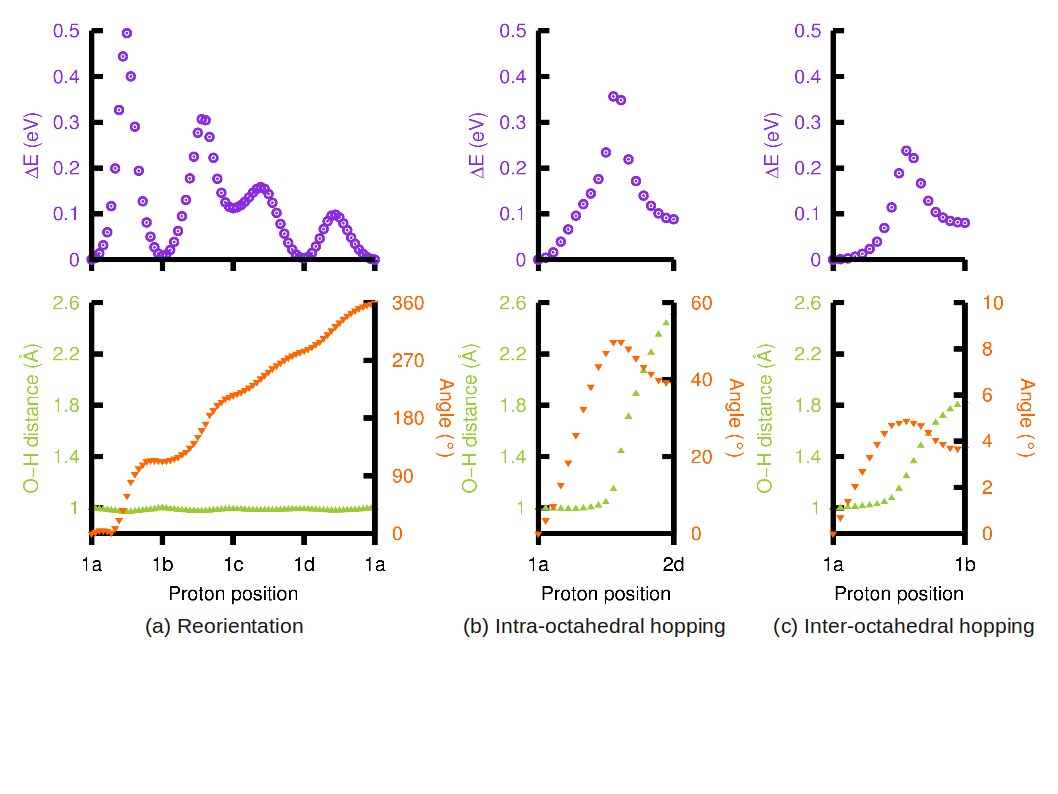}
\caption{Energy profiles and evolution of some geometric quantities along typical Minimum Energy Paths. The angle $\phi$ is between the initial and current O-H direction.}
\label{profile}
\end{figure*} 
\end{center}

Fig.~\ref{profile} illustrates the energy profile for each of the three possible kinds of mechanisms (note this is not an exhaustive list of all possible profiles): 
Fig.~\ref{profile}a shows the energy profile, as well as the evolution of the O-H distance and the angle $\phi$ from the initial O-H direction in the case of a complete turn around an oxygen O1 near the dopant. Using the notations of Tab.~\ref{barrier-value}, it corresponds to the 4 reorientation mechanisms: $1an\Rightarrow1bn\Rightarrow1cn\Rightarrow1dn\Rightarrow1an$. These 4 reorientation barriers have not the same profile at all: not only the barrier height can differ by a factor 5, but also the angle between two stable sites varies from 60\textdegree{} to 120\textdegree{} instead of being set to 90\textdegree{} (case of an ideal cubic perovskite). Figs.~\ref{profile}b and \ref{profile}c give similar information but for intra-octahedral and inter-octahedral hoppings respectively.
Both mechanisms seem to occur in two steps: first a reorientation, slight for inter-octahedral hopping ($\approx$ 5\textdegree) and larger for intra-octahedral hopping ($\approx$ 45\textdegree) in order to get O-H-O aligned, then the jump between both oxygen atoms. This reorientation can be related to what we mentioned as ``flip'' in the previous section.

\section{Discussion}


\subsection{Comparison between Gd-doped BaCeO$_3$ and In-doped CaZrO$_3$}
The present results on Gd-doped BaCeO$_3$ can be compared with previous values computed in In-doped CaZrO$_3$~\cite{Islam2001,Bilic2007,Bilic2008}, as both materials exhibit the same kind of structural distortion: BaCeO$_3$ and CaZrO$_3$ have the same perovskite structure with very close Goldschmidt's tolerance factor (0.94 and 0.92 respectively) and thus have the same orthorhombic structure with P$nma$ space group. However, according to its bigger tolerance factor, BaCeO$_3$ should be slightly less distorted from the cubic structure and thus inter-octahedral transfer may be harder than in CaZrO$_3$. Tab.~\ref{structural-bco-vs-czo} confirms that BaCeO$_3$ is a bit closer to an ideal cubic structure than CaZrO$_3$.

\begin{table}
 \centering
\begin{tabular}{cccc}
& BaCeO$_3$ & CaZrO$_3$[\onlinecite{Bilic2007}] & cubic \\
 a$_c$ (\AA)& 4.44 & 4.06 & --\\
a/a$_c$& 1.41 & 1.39 & 1.41 \\
b/a$_c$& 1.42 & 1.44 & 1.41 \\
c/a$_c$& 2.00 & 2.00 & 2.00 \\
$\overline{\text{A-O}}$/a$_c$ ($\pm\sigma$)& 0.71 ($\pm 0.21$) & 0.72 ($\pm 0.22$)& 0.71 \\
$\overline{\text{B-O}}$/a$_c$ ($\pm\sigma$)& 0.51 ($\pm 0.00$)& 0.52 ($\pm 0.00$) & 0.50\\
A-O-A (deg) & 153.85 & 144.74 & 180.00 \\
B-O-B (deg) & 156.45 & 145.49 & 180.00 \\
\end{tabular}
 \caption{Structural parameters (lattice parameters, cation-oxygen distances and angles) for BaCeO$_3$, CaZrO$_3$ and a fictitious cubic perovskite.}
 \label{structural-bco-vs-czo}
\end{table}

The same tendency is indeed observed with a very large range of possible values for energy barriers from a few 0.01 eV up to nearly 1~eV. 
For instance, in BCGO, reorientation barriers can take a wide range of different values, starting at less than 0.1~eV for barrier between c-type and d-type sites up to 0.5~eV for barrier between a-type and b-type sites. The same results have been found for In-doped CaZrO$_3$~\cite{Bilic2007} except for the fact that the largest barrier can go up to 0.9~eV.

The very small barrier between c and d sites might explain why position 1c is not considered at all in the work of Bilic and Gale~\cite{Bilic2007} (only 7 different positions instead of our 8 positions near a specific B-atom) and 2c near some specific oxygen atoms O2. According to Tab.~\ref{energy-proton}, 1c and 2c are much higher in energy than nearby positions, that is why the reorientation barriers from c-type site are really small.

In both materials, possible inter-octahedral hoppings have a smaller energy barrier than intra-octahedral hopping. This follows from the ability of any oxygen octahedron to bend towards another in the orthorhombic $Pnma$ structure, so that two facing oxygens (belonging to different octahedra) can be made very close to each other. But each octahedron remains rigid, so that its own oxygen atoms cannot be made closer to each other (though a little distortion during the transfer is observed, in agreement with previous calculations~\cite{Cherry1995}). Of course, the inter-octahedral hoppings are possible only when the oxygen atoms involved are close to each other (this corresponds to a/b type within our notation). The c/d type oxygens, which are made further from each other as a result of the tilting process, are excluded from the inter-octahedral motions.

According to those common tendencies, we can suggest that all orthorhombic perovskites behave alike and make some assumptions:
\begin{enumerate}[i/]
 \item rather low barriers ($\lesssim$ 0.2~eV) for inter-octahedral hopping depending on the level of distortion (barrier is smaller as distortion increases)
 \item higher barriers ($\sim$ 0.3-0.6~eV) for intra-octahedral hopping
 \item a wide range of values for reorientation, from less than 0.1~eV up to 0.8~eV, depending on the type of protonic site.
\end{enumerate}

Finally, there is a quantitative difference between both materials concerning the attractive power of the dopant: it seems much harder to escape from Indium in CaZrO$_3$ than from Gd in BaCeO$_3$. The barrier to escape from Indium is on average three times higher than the backward barrier, while in BaCeO$_3$ the escaping barrier is higher only by 50\%. This may be due to the nature of the dopant as suggested by Bjorketun et al.~\cite{Bjorketun2007}, which have shown that energy barriers for proton migration near a dopant can be strongly dependent of its nature. Therefore Gadolinium seems to be a good candidate as a dopant since its power of attraction is low enough to let the proton escape relatively easily.

\subsection{Rate-limiting events}

The rate-limiting process in such distorted system is not so obvious. Contrary to what can be expected, the reorientation is not necessarily much faster than the hopping. Munch and co-workers have found that the proton transfer step is indeed rate limiting in BaCeO$_3$ but of the same order of magnitude as reorientation for SrCeO$_3$~\cite{Munch1999}. More precisely, they computed an activation energy for rotational diffusion in BaCeO$_3$ of 0.07 eV for O$_1$ and 0.11 eV for O$_2$, close to the values we get for the lowest reorientation barriers. In earlier work~\cite{Munch1997}, they found for Ba\{Ce,Zr,Ti\}O$_3$ that reorientation happens much faster with a time scale of $\sim 10^{-12}$~s, while proton transfer occurs at a time scale of $10^{-9}$~s. However the three materials have been studied in their cubic structure, thus preventing the low-barrier inter-octahedral transfer. Gomez and co-workers~\cite{Gomez2007} precise that the rate-limiting process in orthorhombic structure is an intra-octahedral transfer. The fact that most of these studies only focus on the cubic structure might explain why the transfer step has been thought to be rate-limiting. 


\section{Conclusion}

In this work, we have performed density-functional calculations on fully hydrated Gd-doped barium cerate and computed in an exhaustive way the Minimum Energy Paths between stable protonic sites close and far from the Gd dopant.

Proton transport in perovskites is usually described as a two-step Grotthuss-type diffusion mechanism: a quick reorientation, followed by a transfer to another oxygen~\cite{Kreuer1999}. However, even if this is correct in principle, we have found that in Gd-doped BaCeO$_3$, the reorientation is not necessarily a fast process compared to transfer. In this distorted perovskite with orthorhombic $Pnma$ space group, inter-octahedral hoppings with rather low barriers $\sim 0.2$~eV do exist.
Also, reorientation mechanisms can be very different from one site to another and thus take a wide range of possible values from 0.02~eV up to 0.54~eV. To a lesser extent, the same argument can be applied to intra-octahedral hopping for which the energy barrier varies between 0.22 and 0.58~eV.

All these results are qualitatively comparable with a previous work focused the orthorhombic perovskite In-doped CaZrO$_3$~\cite{Bilic2007}. The low barriers found for inter-octahedral hopping in these orthorhombic structures suggest that protonic diffusion could be much faster in such structure than in the cubic one, since an inter-octahedral hopping is equivalent to two intra-octahedral transfers but with a higher rate. All the barrier values will be exploited in Kinetic Monte-Carlo simulations to check the actual rate of reorientation versus hopping, and simulate proton trajectories on larger space and time scales.

Finally, gadolinium in barium cerate seems to be interesting as a dopant as it acts like a shallow trap for protons, with rather low escaping barrier (compared to indium in calcium zirconate), enabling the proton to diffuse quite easily. However, other trivalent dopants could be tested to check whether they have better properties for protonic diffusion.

\section{Acknowledgements}
This work was performed using the HPC resources of the TERA-100 supercomputer of CEA/DAM and from GENCI-CCRT/CINES (Grants 2010-096468 and 2011-096468).
We acknowledge that some contributions to the present work have been achieved using the PRACE Research Infrastructure resource (machine CURIE) based in France at Bruy\`eres-le-Chatel (Preparatory Access 2010PA0397).

\section{Appendix: computation of energy barriers far from dopants using a charged supercell}
The barriers corresponding to motions far from the dopant, {\it i.e.} from a ``f'' configuration to another ``f'' configuration, have been recomputed using an undoped supercell in which the charge of the proton is compensated by a uniform charged background (jellium), as frequently done for the simulation of charged defects. In such cases, there are 16 different motions: 8 reorientations, 5 intra-octahedral hoppings and 3 inter-octahedral hoppings. This corresponds to 30 barrier values. The energy barriers obtained using this method are compared to the ones obtained using the doped supercell in Tab.~\ref{barrier-bco}: the values obtained using the two methods are the same within 0.01 eV, confirming that a 80-atom supercell is large enough to contain a region ``close'' to the dopant and a region ``far'' from it. The proton ``far'' from the dopant does not feel the influence of Gd atoms, and can be considered as in pure BaCeO$_3$. Besides, the fact that we get the same values in both cases suggests that the jellium only induces a systematic shift in total energies, but does not affect energy differences.

\begin{table}
 \centering
 \begin{tabular}{ccccc}
Barrier & \multicolumn{2}{c}{pure BaCeO$_3$} & \multicolumn{2}{c}{``far'' BaCeGdO$_3$} \\
Reorientation & $\rightarrow$ & $\leftarrow$ & $\rightarrow$ & $\leftarrow$ \\
1a-1b & 0.54 & 0.54 & 0.54 & 0.54 \\
1b-1c & 0.33 & 0.18 & 0.33 & 0.18 \\
1c-1d & 0.06 & 0.15 & 0.06 & 0.15 \\ 
1d-1a & 0.09 & 0.14 & 0.08 & 0.14 \\
2a-2b & 0.36 & 0.49 & 0.36 & 0.49 \\
2b-2c & 0.33 & 0.17 & 0.33 & 0.17 \\
2c-2d & 0.03 & 0.08 & 0.02 & 0.08 \\
2d-2a & 0.20 & 0.17 & 0.20 & 0.17 \\
Hopping & $\rightarrow$ & $\leftarrow$ & $\rightarrow$ & $\leftarrow$ \\
1a-2d & 0.50 & 0.37 & 0.50 & 0.37 \\
1b-2d & 0.45 & 0.31 & 0.45 & 0.31 \\
1c-2b & 0.36 & 0.47 & 0.36 & 0.47 \\
1d-2b & 0.42 & 0.44 & 0.42 & 0.44 \\
2a-2c & 0.39 & 0.36 & 0.39 & 0.36 \\
1a-1b & 0.19 & 0.20 & 0.19 & 0.20 \\
2a-2a & 0.21 & -- & 0.21 & -- \\
2b-2b & 0.16 & -- & 0.16 & -- \\
 \end{tabular}
 \caption{Comparison of barrier values ``far'' from the dopant in BCGO and in pure charged BaCeO$_3$.}
 \label{barrier-bco}
\end{table}


\bibliography{article}

\end{document}